# Fast Fourier-transform calculation of artificial night sky brightness maps

Salvador Bará,[1,*] Fabio Falchi,[2] Riccardo Furgoni,[2] and Raul C. Lima[3,4]

[1] Departamento de Física Aplicada, Universidade de Santiago de Compostela, 15782 Santiago de Compostela, Galicia.
[2] Istituto di Scienza e Tecnologia dell'Inquinamento Luminoso (Light Pollution Science and Technology Institute), 36016 Thiene, Italy.
[3] Physics, Escola Superior de Saúde, Politécnico do Porto, Portugal.
[4] CITEUC – Centre for Earth and Space Research, University of Coimbra, Portugal.
[*] salva.bara@usc.es

**Abstract**

Light pollution poses a growing threat to optical astronomy, in addition to its detrimental impacts on the natural environment, the intangible heritage of humankind related to the contemplation of the starry sky and, potentially, on human health. The computation of maps showing the spatial distribution of several light pollution related functions (e.g. the anthropogenic zenithal night sky brightness, or the average brightness of the celestial hemisphere) is a key tool for light pollution monitoring and control, providing the scientific rationale for the adoption of informed decisions on public lighting and astronomical site preservation. The calculation of such maps from satellite radiance data for wide regions of the planet with sub-kilometric spatial resolution often implies a huge amount of basic pixel operations, requiring in many cases extremely large computation times. In this paper we show that, using adequate geographical projections, a wide set of light pollution map calculations can be reframed in terms of two-dimensional convolutions that can be easily evaluated using conventional fast Fourier-transform (FFT) algorithms, with typical computation times smaller than $10^{-6}$ s per output pixel.

Keywords: Light pollution ; Atmospheric optics ; Photometry ; Radiometry ; Fourier transforms.

## 1. Introduction

The sustained increase of the anthropogenic light emissions poses a relevant threat to the performance of optical astronomical observatories in many regions of the world [1-2]. Artificial light scattered by the atmosphere severely reduces the contrast of the objects in science images, imposing stringent requirements on the dynamic range and the resolution of the imaging and spectrometric detectors used for their study. According to recent reports, during the period 2012-2016 artificial light emissions grew worldwide at an average rate of about 2% per year, both in total radiance and artificially lit area [3]. Growth rates of 0-20% per year have been documented in other periods of time for different regions of the world, revealing a highly inhomogeneous distribution across countries [4]. Besides its detrimental effects on astrophysical observations, light pollution is recognized nowadays as a global issue whose negative consequences impact in unintended ways the natural environment, accelerate the loss of sky-related intangible cultural heritage and, according to recent findings, can potentially affect relevant aspects of human health [5-16].

A significant effort has been devoted in the last years to the development and validation of theoretical models describing the propagation of artificial light through the atmosphere [17-31]. Their common goal is to determine the value of different magnitudes of interest (e.g. the anthropogenic zenithal night sky brightness) at any desired observing site, in terms of the radiance distribution of the surrounding artificial light sources, the optical properties of the atmosphere, the characteristics of the built spaces and the intervening terrain, and the presence of obstacles that could block or otherwise modify the free propagation of radiance. Light pollution propagation models usually calculate these magnitudes of interest as integral transforms (weighted integrals) of the spatial, angular, and spectral radiance distribution of the artificial light sources located within the region that effectively influences the observing site. Depending on the particular conditions of the sources and the atmosphere, this region may span a radius up to a few hundred kilometers: light pollution effects can be recorded at great distances from the cities that produce them. Theoretical models differ from each other in the particular assumptions and approximations used to build the kernel of the transform (e.g. the presence or not of obstacles, single versus multiple scattering, particular expressions for the molecular and aerosol concentration profiles, assumed phase scattering function, etc). Ground radiance data with medium to high spatial resolution and nearly worldwide coverage are available, among other data sources, from the legacy archives of the Defense Meteorological Satellite Program Operational Linescan System (DMSP-OLS) [20, 32-33], the current Suomi-NPP VIIRS-DNB datasets [34-38], and the International Space Station (ISS) DSLR images [39-41], the first two panchromatic (0.5-0.9 µm band) and the latter trichromatic RGB.

Calculating the light pollution magnitudes of interest for any given observing site from satellite radiance images involves the need of performing multiple sums over pixels. If these magnitudes are to be determined for a country-wide region or for the whole planet with sub-kilometric spatial resolution the computational load increases vastly. In this work we take advantage of the fact that for an important subset of light propagation kernels, namely those that can be considered shift-invariant within the region of interest in an appropriate reference frame, the light pollution propagation integrals can be rewritten as two-dimensional convolutions, and their evaluation can be carried out very efficiently using standard techniques in the Fourier domain. The computational gains come from the fact that a convolution integral in the direct space is strictly equivalent to a pixel-wise multiplication in the Fourier-transformed one, and that extremely efficient algorithms for performing the direct and inverse Fourier transformations required to apply this method are standard features of almost every scientific programming environment and are even available in widely distributed free GIS applications.

In this paper we develop this approach and present an example of how Fast Fourier transform algorithms (FFT) can be applied to the efficient computation of zenithal sky brightness and average hemispheric sky brightness maps across extended patches of territory (with areas of order $10^6$ km$^2$) with typical calculation times below $10^{-6}$ s per output pixel, for the particular combination of hardware and software used in our work. To that end, in section 2 we briefly revisit the basics of the Fourier convolution theorem, as well as the conditions under which it can be applied to the present issue. Section 3 presents a practical example of application. The possibilities and limitations of this method are addressed in the Discussion (section 4), and conclusions are drawn in section 5.

## 2. Methods

### 2.1. Light pollution functions and integral transforms

Let us denote by $B(\boldsymbol{r})$ any generic radiant or photometric magnitude relevant for light-pollution studies, being $\boldsymbol{r}$ the position vector of the observing site. In the present context $B(\boldsymbol{r})$ may be any member of a wide set of functions linearly related to the source radiance, e.g., the zenithal night sky brightness, the brightness in any other direction of the upper hemisphere, the average sky radiance, or the horizontal illuminance, among others [42], including, where appropriate, their spectral density distributions. Let us further denote by $L(\boldsymbol{r}', \boldsymbol{\alpha}'; \lambda)$ the spectral radiance emitted in the direction described by the two-dimensional vector $\boldsymbol{\alpha}' = (z', \varphi')$ by a source located at $\boldsymbol{r}'$, being $z'$ and $\varphi'$ the zenith angle and the azimuth, respectively, in the source reference frame, and $\lambda$ the wavelength. Since the actual irradiances associated with streetlight sources are very far from the thresholds required to produce any kind of non-linear effects, the light pollution propagation through the atmosphere takes place in the linear regime and the value of $B(\boldsymbol{r})$ can be obtained by adding up the contributions of all individual sources as:

$$B(\boldsymbol{r}) = \int_\Lambda \int_\Omega \int_{S'} \int_{\Omega'} G(\boldsymbol{r}, \boldsymbol{\alpha}; \boldsymbol{r}', \boldsymbol{\alpha}'; \lambda) L(\boldsymbol{r}', \boldsymbol{\alpha}'; \lambda) \mathrm{d}^2\boldsymbol{\alpha}' \, \mathrm{d}^2\boldsymbol{r}' \mathrm{d}^2\boldsymbol{\alpha} \, \mathrm{d}\lambda \, , \qquad (1)$$

where $\boldsymbol{\alpha} = (z, \varphi)$ is the two-dimensional direction vector in the observer reference frame, and $G(\boldsymbol{r}, \boldsymbol{\alpha}; \boldsymbol{r}', \boldsymbol{\alpha}'; \lambda)$, the kernel of this integral transform, is the function describing the elementary contribution of a unit amplitude (in the Dirac-delta sense) spectral radiance source to the final value of $B(\boldsymbol{r})$. $\mathrm{d}^2\boldsymbol{\alpha}$, $\mathrm{d}^2\boldsymbol{\alpha}'$, and $\mathrm{d}^2\boldsymbol{r}'$ are short-hand notations for the infinitesimal elements of solid angle ($\mathrm{d}^2\boldsymbol{\alpha} = \sin z \, \mathrm{d}z \mathrm{d}\varphi$, $\mathrm{d}^2\boldsymbol{\alpha}' = \sin z' \, \mathrm{d}z' \mathrm{d}\varphi'$, in spherical coordinates) and surface ($\mathrm{d}^2\boldsymbol{r}' = \mathrm{d}x' \mathrm{d}y'$, in Cartesian ones), respectively, and $d\lambda$ is the elementary spectral interval. The integrations are carried out over the surface $S'$ of the territory encompassing the intervening sources, the $2\pi$ sr set of directions contained in the upper hemisphere of the sources ($\Omega'$), the upper hemisphere ($\Omega$) of the observer when appropriate (e.g. for computing the horizontal irradiance) and the relevant spectral interval ($\Lambda$). The determination of the particular form of $G(\boldsymbol{r}, \boldsymbol{\alpha}; \boldsymbol{r}', \boldsymbol{\alpha}'; \lambda)$, either analytically or by numerical computation, is the main task that has to undertake a light propagation models developer. A discrete version of equation (1) suitable for working with pixelated radiance data from satellite imagery, is

$$B(\boldsymbol{r}_i) = \sum_{q=1}^{Q}\sum_{j=1}^{J}\sum_{k=1}^{K}\sum_{l=1}^{L} G(\boldsymbol{r}_i, \boldsymbol{\alpha}_j; \boldsymbol{r}'_k, \boldsymbol{\alpha}'_l; \lambda_q) L(\boldsymbol{r}'_k, \boldsymbol{\alpha}'_l; \lambda_q)\, \Delta^2 \boldsymbol{\alpha}'_l\, \Delta^2 \boldsymbol{r}'_k \Delta^2 \boldsymbol{\alpha}_j\, \Delta\lambda_q \,. \quad (2)$$

## 2.2. Factorable sources and factorable regions

Factorable sources are light sources whose radiance can be factored out into a spatially dependent term and an angular-wavelength one [43]. This concept can be applied to the radiance of wide spatial regions, considering them factorable if the following equality holds:

$$L(\boldsymbol{r}', \boldsymbol{\alpha}'; \lambda) = L_1(\boldsymbol{r}') L_2(\boldsymbol{\alpha}', \lambda)\,, \quad (3)$$

that is, if the shape of the spectral and angular radiance pattern, $L_2(\boldsymbol{\alpha}', \lambda)$, is essentially the same for all points of the territory under study, and the only variation between points is their overall amount of emissions, $L_1(\boldsymbol{r}')$. This condition is implicitly used in the calculation of many light pollution maps from pixelated satellite radiance data, under the simplifying assumption that each pixel radiates light in approximately the same way, only differing in their overall flux, which is in turn estimated from the radiance recorded by the on-orbit radiometer. This assumption can be deemed reasonable for many cases of interest, based on the within-pixel averaging of the angular and spectral emission of the individual light sources contained in medium-sized ground pixels (~ hundreds of m wide) and the subsequent addition of the contributions of many different pixels to obtain the final result in equation (2). Henceforth, for the sake of simplicity, we will assume that the sources are factorable in strict sense, as described by equation (3). However, our results are straightforwardly extensible to more general situations. A condition substantially less stringent than (3) is the one in which the light sources can be grouped into $t=1,...,T$ different classes of factorable ones, such that the pixel radiances can be written as:

$$L(\boldsymbol{r}',\boldsymbol{\alpha}';\lambda) = \sum_{t=1}^{T} L_{1t}(\boldsymbol{r}')L_{2t}(\boldsymbol{\alpha}',\lambda). \quad (4)$$

The application of our results to the situation described by equation (4) is immediate.

Under condition (3), equation (1) becomes the simpler transform:

$$B(\boldsymbol{r}) = \int_{S'} K(\boldsymbol{r},\boldsymbol{r}')L_1(\boldsymbol{r}')\mathrm{d}^2\boldsymbol{r}' , \quad (5)$$

whose kernel (which in linear system analysis is also known as the *point spread function*, or PSF) is given by:

$$K(\boldsymbol{r},\boldsymbol{r}') = \int_\Lambda \int_\Omega \int_{\Omega'} G(\boldsymbol{r},\boldsymbol{\alpha}; \boldsymbol{r}',\boldsymbol{\alpha}';\lambda)L_2(\boldsymbol{\alpha}',\lambda)\mathrm{d}^2\boldsymbol{\alpha}' \, \mathrm{d}^2\boldsymbol{\alpha} \, \mathrm{d}\lambda , \quad (6)$$

**2.3. Shift-invariant kernels and Fourier evaluation of convolution integrals**

Equation (5) is a superposition integral widely used in light pollution propagation models. An additional assumption, commonly adopted in a wide subset of models in order to simplify the analytical calculation of the PSF $K(\boldsymbol{r},\boldsymbol{r}')$, is the presence of a layered atmosphere, whose properties (molecular and aerosol concentration) vary along the altitude axis but are constant in the horizontal directions. In that case, neglecting the effect of local obstacles and the different altitudes above sea level of the sources and the observing points, the PSF turns out to be transversally shift-invariant, i.e. it only depends on the relative position of the observing site with respect to the source, but not on the absolute position of each. We have then $K(\boldsymbol{r},\boldsymbol{r}') = K(\boldsymbol{r} - \boldsymbol{r}')$, and the integral transform (5) becomes a two-dimensional convolution of the form

$$B(\boldsymbol{r}) = \int_{S'} K(\boldsymbol{r} - \boldsymbol{r}')L_1(\boldsymbol{r}')\mathrm{d}^2\boldsymbol{r}' , \quad (7)$$

The usefulness of this approach is that convolution integrals can be evaluated either in the direct space, by sequentially performing for each $r$ the summation over all $r'$ as indicated by equation (7), or in the transformed Fourier domain, where this operation becomes a simple product of functions. Let us recall that under very general conditions (e.g. being absolutely integrable, which is a trivial feature of satellite radiance images), the Fourier transform (or *Fourier spectrum*) $F(\boldsymbol{v})$ of a two-dimensional function $f(\boldsymbol{r})$ is given by [44-45]:

$$F(\boldsymbol{v}) = \int_{\infty} f(\boldsymbol{r}) \exp(-i2\pi \, \boldsymbol{v} \cdot \boldsymbol{r}) \mathrm{d}^2 \boldsymbol{r}, \qquad (8)$$

where $\boldsymbol{v} = (v_x, v_y)$ is a vector whose components play the role of spatial frequencies (units m$^{-1}$) along the two orthogonal dimensions of the inverse space domain. The Fourier transform is a useful tool to address several light pollution issues (see, e.g. [46]). A relevant property of the Fourier transform, the so-called convolution theorem, states that convolution integrals like equation (7) become simple products in the Fourier domain [45]. That is, denoting by $\widetilde{B}(\boldsymbol{v})$, $\widetilde{K}(\boldsymbol{v})$, and $\widetilde{L}(\boldsymbol{v})$ the Fourier transforms of $B(\boldsymbol{r})$, $K(\boldsymbol{r})$, and $L_1(\boldsymbol{r})$, respectively, the following equality holds:

$$\widetilde{B}(\boldsymbol{v}) = \widetilde{K}(\boldsymbol{v})\widetilde{L}(\boldsymbol{v}). \qquad (9)$$

The convolution theorem provides an alternate pathway for obtaining $B(\boldsymbol{r})$, consisting of calculating first the Fourier transforms of $K(\boldsymbol{r})$, and $L_1(\boldsymbol{r})$, multiplying them to get $\widetilde{B}(\boldsymbol{v})$ as per equation (9), and applying an inverse Fourier transform to $\widetilde{B}(\boldsymbol{v})$ to get the desired $B(\boldsymbol{r})$ back in the spatial domain. The practical relevance of this theorem for the calculation of light pollution maps stems from the fact that very efficient numerical algorithms are available for computing the discrete version of equation (8), i.e. the *Discrete Fourier Transform* (*DFT*) $F(k,l)$ of a $N \times M$ matrix $f(n,m)$, which is defined as:

$$F(k,l) = \sum_{n=1}^{N}\sum_{m=1}^{M} f(n,m) \exp\left\{-i2\pi\left[\frac{(k-1)(n-1)}{N} + \frac{(l-1)(m-1)}{M}\right]\right\}. \qquad (10)$$

These algorithms, highly optimized since the pioneering work of Cooley and Tukey [47], and collectively known as *Fast Fourier Transforms* (*FFT*), allow to reduce significantly the number of basic operations required to compute the convolution of two matrices. If the sizes of the matrices $L_1(\boldsymbol{r})$ and $K(\boldsymbol{r})$ are $N_L \times N_L$ and $N_K \times N_K$, respectively, a conventional discrete convolution algorithm requires computing times of order $O(N_L{}^2 N_K{}^2)$, whereas *FFT* algorithms allow to obtain the same result in times of order $O[(N_L + N_k)^2 \log_2(N_L + N_k)^2]$ [48]. The time savings soon become relevant as the matrix sizes increase: for instance, for a 1000 km x 1000 km satellite radiance tile with 0.5 km resolution and a PSF matrix 300 km x 300 km wide with the same spatial resolution, the basic time required to perform their convolution using *FFT* is about $10^4$ times smaller than the one required to perform it directly. Hence, for medium to large-sized matrices, the numerical calculation of a convolution can be accomplished substantially faster by following the apparently more involved route of *FFT* transforming the matrices into the Fourier domain, performing a pixel-wise multiplication of their Fourier spectra, and applying an inverse *FFT* (*iFFT*) afterwards.

## 2.4. Reprojecting satellite radiance maps

The formulation of the light pollution propagation as a convolution integral requires operating in a reference frame with uniform scale throughout the whole area under study, in order to preserve the PSF shift-invariance. Several widely used satellite radiance products, however, are provided in geographical projections whose spatial scale of length is widely variable, depending on the region of the map. The VIIRS Day/Night Band Nighttime Lights monthly or annual composites, for instance, provide the radiance data in a uniform WGS84 longitude-latitude grid or *plate carrée* [49] with 15 arc-second resolution [38, 50]. Each 15x15 arc-second$^2$ pixel of the composites, then, has a different width (in km) along the longitude axis depending on its precise latitude and, consequently, also a different area in squared length units.

Although no geometrical projection can map globally a sphere into a plane preserving exactly all the distances between any possible pairs of points, several map projections provide enough accurate approximations to the uniform scale condition over regions of the Earth of reasonable extent. One of them is the widely used Universal Transverse Mercator (UTM) [49, 51], which was adopted as the base of the official cartography of many world countries and organizations. The UTM is a cylindrical conformal projection that divides the Earth in 60 longitude regions, each of them 6 degrees wide, whose corresponding plane maps are obtained by projecting the surface of the planet onto a cylinder whose line of tangency to the sphere is the central meridian of the region. The scale factor for any point of the map only depends on its distance to the central meridian. The scale factor along the central meridian, $k_0$, can be made equal to 1 or be deliberately reduced to a slightly smaller value, in order to keep the average scale within appropriate limits across the whole map region [51]. In the latter case, the projection cylinder has a slightly smaller radius and, instead of being strictly tangent to the sphere, intersects it at short distances from the central meridian.

The scale factor for a point of coordinates $(x, y)$ in a UTM map, from equations 8-2 and 8-4 of Snyder [51], is given by

$$k(x,y) = k_0 \cosh\left[\frac{x}{k_0 R}\right], \qquad (11)$$

where $R$ is the Earth radius and $k_0$ is the central scale, usually chosen as $k_0 = 0.9996$. This choice provides $k = 1$ at $x = \pm 180$ km from the central meridian. Being the UTM a conformal projection, the length scale is the same along both coordinate axes and the area element scales as $k^2$.

The scale distortions introduced by the UTM projection for reasonably sized light pollution maps are relatively small. If the desired map region spans an area of 1000 x 1000 km² (i.e. 500 km either side from the central meridian), and allowing e.g. 300 additional km to correctly account for the contributions of the radiance sources located outside the limits of the map to the sky brightness of the map rim regions, the scale at the outermost limits of the required satellite radiance tile ($x = \pm 800$ km) is, according to equation (11) for $R = 6367$ km, $k = 1.0075$, i.e. less than 1% error. The surface scale error is, for these limiting source regions located 300 km outside the map, barely 1.5%.

The reprojection of the WGS84 uniformly spaced lat-lon tiles of the VIIRS-DNB onto suitable planar grids (e.g. the ETRS89 UTM zone system) can be easily performed using any available GIS software package commonly used in geospatial analysis, including their free versions (e.g. QGIS). Since the pixel grid mosaic resulting from the reprojection is not strictly coincident with the original one, some interpolation procedure must be used to assign the projected pixel radiance values. Nighttime lights images of the Earth represent an intermediate situation between continuous spatial functions and discrete areas. Overall, nearest neighbor interpolation provides a reasonable trade-off for the estimation of the pixel radiance in reprojected maps.

**2.5. Overall workflow**

The steps to calculate light pollution maps using the Fourier transform approach are schematically depicted in Figure 1. Note that since the convolution obtained using *FFT* is a circular (rather than a linear) one, the matrices $L_1(\boldsymbol{r})$ and $K(\boldsymbol{r})$ shall be padded with zeros until reaching a common size of $N_L + N_k - 1$ rows and columns before calculating their Fourier transforms (the generalization for non-square matrices is immediate).

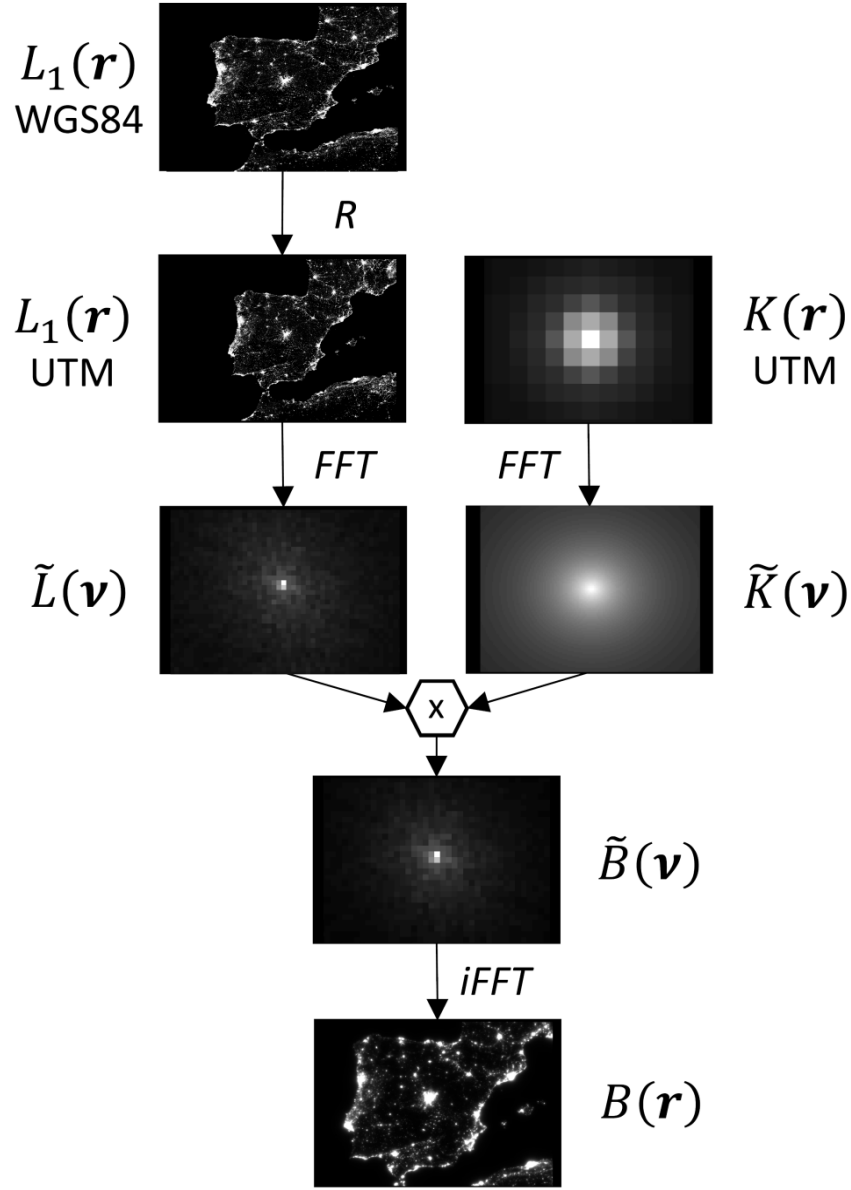

**Figure 1:** Steps for calculating light pollution maps using the Fourier convolution theorem. *R*: map reprojection; *FFT*: Fourier-transform; *iFFT*: inverse Fourier-transform. $\tilde{B}(\boldsymbol{v})$, $\tilde{K}(\boldsymbol{v})$, and $\tilde{L}(\boldsymbol{v})$ are in general complex functions; their absolute values are plotted in this figure.

## 3. Results

As an example of application, we show in this section the results of the calculation of two different types of light pollution maps corresponding, respectively, to the spatial distribution of the zenithal night sky brightness and to the all-sky light pollution ratio, according to the PSFs described by Cinzano and Falchi [22], and Duriscoe et al. [52], respectively.

The $L_1(r)$ radiance map providing information on the artificial light sources corresponds to the Iberian Peninsula section (Figure 2) of an ample region of West Europe and North Africa extracted from the VIIRS stable lights 2016 composite tile 2 version "vcm-orm-ntl" (VIIRS Cloud Mask - Outlier Removed - Nighttime Lights) [38], reprojected to ETS89 UTM zone 30N (EPSG 25830) using nearest-neighbor interpolation, with square output pixels of uniform size 404.4 m.

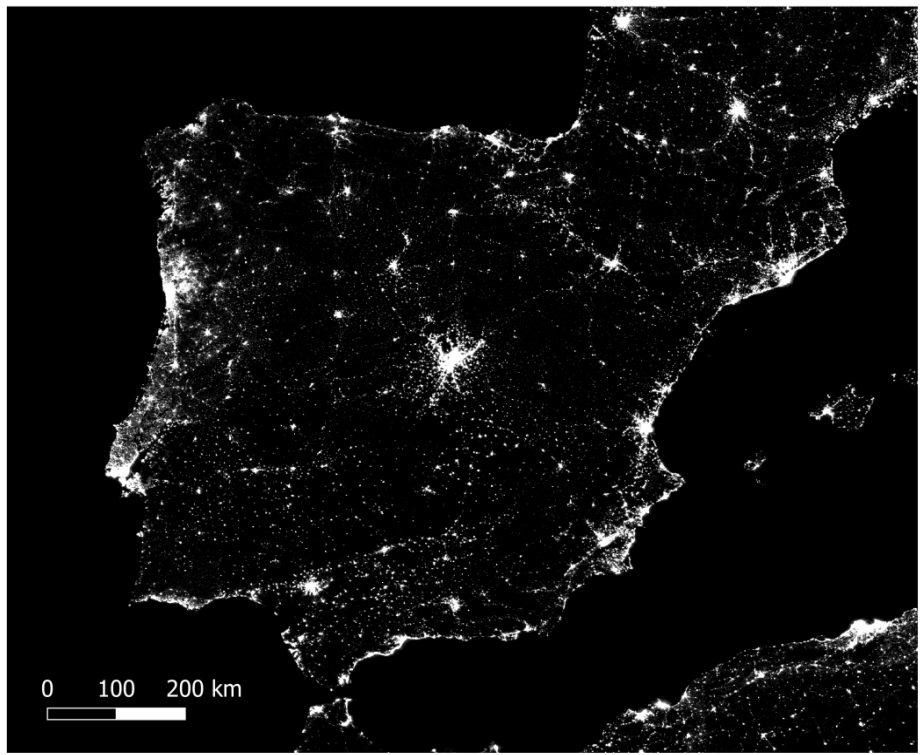

**Figure 2:** Radiance data from the VIIRS stable lights 2016 composite, version "vcm-orm-ntl" (VIIRS Cloud Mask - Outlier Removed - Nighttime Lights) [38], reprojected to ETS89 UTM zone 30N (EPSG 25830) using nearest-neighbor interpolation.

For the calculation of the zenithal night sky brightness (ZSB), we used the LPTRAN PSF described by Cinzano and Falchi [22] (see Fig. 1 in that reference) for a layered atmosphere with clarity K=1 (visibility 26 km). Radiant sources extending up to 195 km beyond the region of interest were included in the analysis. The resulting map of the zenithal night sky brightness distribution is shown in Figure 3.

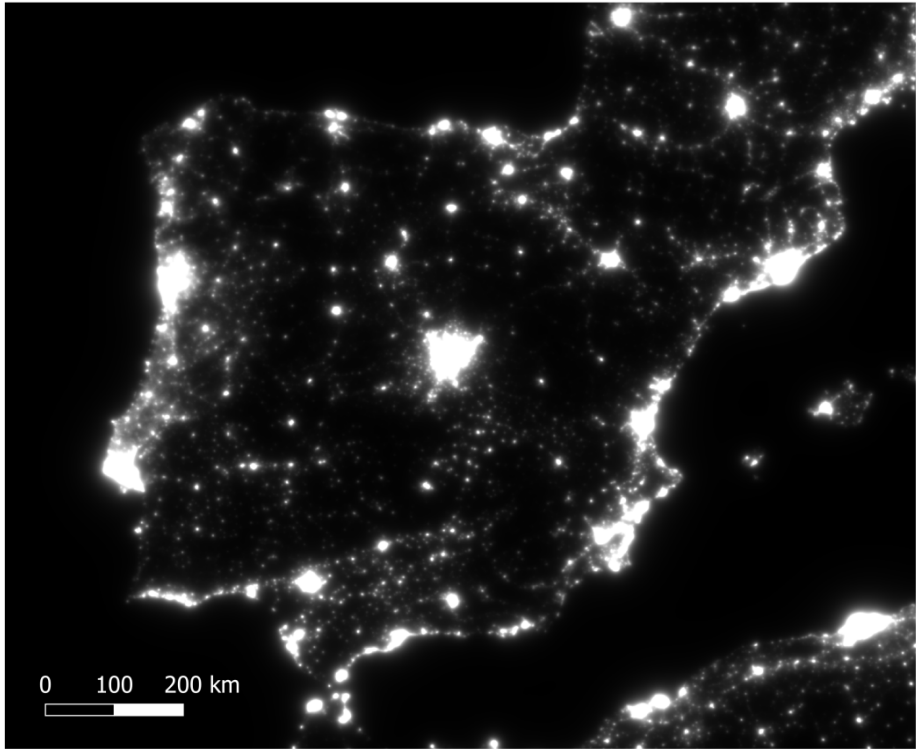

**Figure 3:** Spatial distribution of the zenithal night sky brightness (ZSB) computed from the sources shown in Figure 2 with an atmospheric clarity parameter K=1 (see text for details), and displayed as a linear grayscale of sky radiance in arbitrary relative units.

The all-sky light pollution ratio (ALR) introduced by Duriscoe et al. [52] is defined as the average, over the celestial hemisphere above the observer, of the artificial sky luminance expressed in units of 250 μcd cm$^{-2}$, a nominal averaged brightness taken as reference for pristine natural skies. According to equations (9) and (11) of [52], the PSF for an atmosphere with a Garstang K=0.35 parameter (visibility 65 km; see reference for details) can be well approximated to distances up to 300 km by the analytical expression:

$$K(\boldsymbol{r}-\boldsymbol{r}') = c \times d[km]^{-\alpha(d)}, \qquad (12)$$

where $d[km] \equiv \|\boldsymbol{r}-\boldsymbol{r}'\|$ is the distance between the source and the observation point, expressed in km. The distance-dependent exponent is given by:

$$\alpha(d) = 2.3\left(\frac{d[km]}{350}\right)^{0.28}, \qquad (13)$$

with $c = 1/562.72$. The resulting ALR map for the sources in Figure 2 computed by means of FFT is shown in Figure 4.

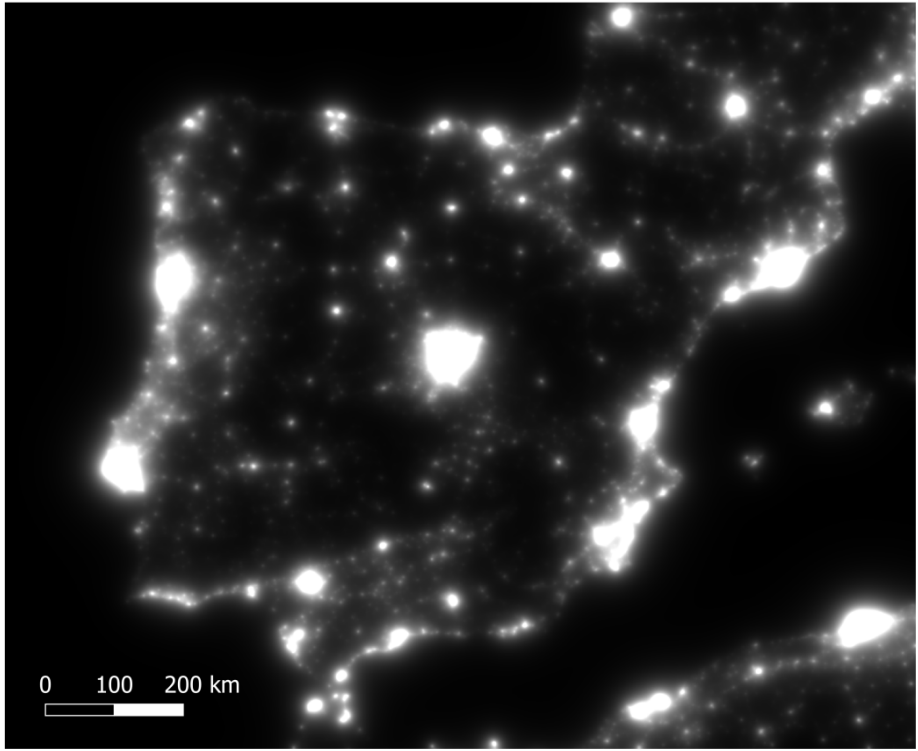

**Figure 4:** Spatial distribution of the all-sky light pollution ratio (ALR) computed from the sources shown in Figure 2 with an atmospheric clarity parameter K=0.35 (see text for details), and displayed as a linear grayscale of hemispherical average sky radiance, in arbitrary relative units.

The full reprojected VIIRS image used in the calculations, as well as the resulting ZSB and ALR maps, significantly wider than the region shown in Figs 2 to 4, were of size 140 Mpixel each (11920 x 11804). The total calculation time (Matlab R2011a running on Intel Core i7, 2.40 GHz, 16 GB RAM), including the time required to save the output files, was 42.86 s and 40.77 s, respectively. This corresponds to calculation times per output pixel of $2.5 \times 10^{-7}$ s and $2.6 \times 10^{-7}$ s, respectively.

## 4. Discussion

The evaluation of light pollution propagation integrals using the Fourier convolution theorem by means of FFT algorithms provides a time-efficient way of computing light pollution maps for wide regions of the Earth with sub-kilometer spatial resolution. The method can be applied as far as the PSF associated with the structure of the sources, the atmospheric conditions, the characteristics of the built spaces and the orographic features of the intervening terrain allow to consider that the radiance distribution of the pixels of the satellite maps is factorable in the sense defined by equations (3) or (4) of section 2.2 and, additionally, that the resulting PSF is shift-invariant.

Note that the FFT approach provides exactly the same results as the conventional calculation of the light pollution maps by means of weighted sums over pixels in the spatial domain, to within the numerical round-off errors provided by the processor used in the calculations. It is important to stress that the FFT algorithm is not an approximation to the exact value of the discrete convolution, but an alternative and fully equivalent way of computing it, taking advantage of the small number of basic operations required to perform the transforms and the fact that a convolution in the spatial domain becomes a pixel-wise multiplication in the spatial frequency domain. Highly efficient, optimized routines for performing Fast-Fourier transforms in one or two dimensions are available in most scientific calculation packages. The interested reader may easily apply them for performing the calculation steps described in Figure 1.

Regarding the limitations of this approach, it shall be kept in mind that the shift-invariance condition (a usual assumption in many light pollution propagation calculations) is only approximately fulfilled in actual situations. Whereas small obstacles and irregularities at the microscale (e.g. in local surface reflectance) can be handled by this model using within-pixel statistical averaging and, if needed, pixel-wise correction of the satellite raw radiance data, other relevant factors like regional orographic features or noticeable altitude differences between the sources and the observers cannot be easily accommodated for. When these factors become relevant, the PSF losses the translational invariance, and the evaluation of the light pollution propagation must be carried out using equation (5) in the spatial domain. Note also that the PSF invariance requires a uniform spatial scale, so it is recommended that the FFT calculations of light pollution maps be made using appropriately sized Earth projection patches. When working with satellite radiance data provided in a uniform longitude-latitude grid (e.g. WGS84) it is advisable to reproject them onto a geographical coordinate grid that preserves the distances to a suffcient degree of approximation within the region of interest (e.g. the appropriate version of the UTM system). The calculations shown as examples in section 3 above were performed using the UTM zone 30N (EPSG 25830) projected grid, which is strictly valid to within the nominal precision described in section 2.4 for the wide area comprised within 34.75° N and 62.33° N latitude and −6.00° W and 0.00° W longitude. In Figures 2 to 4 this corresponds to a North-South band centered on the middle Iberian Peninsula meridian and approximately of that width. Some distortion in the spatial scale is expected to appear outside this zone, progressively increasing as we travel farther away to the East or West from it, but that effect can be remediated by computing the light pollution maps for these peripheral zones using the adequate projection system for each longitude band. Given that FFT calculation time is not a strong constraint, light pollution maps for extended regions of the planet can be calculated using the appropriate UTM projection for each longitude band. A potentially interesting alternative possibility, not developed here, would be to perform the calculations in the native VIIRS longitude-latitude reference system (WGS84), using a slowly varying PSF defined on that grid and dependent on the latitude. This PSF can be considered constant, up to a given accuracy, within sufficiently narrow latitude intervals. The calculation of the light pollution maps would then be carried-out using narrow latitude band regions instead of narrow longitude ones, as done here using the UTM reprojection approach.

The spatial detail with which the territorial distribution of the light sources can be included in the calculations is limited by the effective pixel size of the presently available radiance datasets. This pixel size varies from less that 1 m for airborne photographs to tens or hundreds of meters for satellite imagery. Satellite images are the primary input for the calculation of light pollution magnitudes at wide territorial scales, since the use of large mosaics of nocturnal aerial images is in many cases impractical. Whereas the detailed distribution of individual streetlights has a negligible effect on the aggregated light pollution observed at large distances from the cities, it can be relevant for adequately describing light pollution effects only observed at the urban microscale, where the non-axisimmetry and variance of the upward radiance pattern, $L_2(\boldsymbol{\alpha'}, \lambda)$ in eq. (3), will play a larger role in introducing errors. It may be anticipated that the improvement in spatial resolution (and also in spectral discrimination capability) of new Earth-orbiting platforms will lead to a corresponding improvement of the estimation of light pollution effects at very short distances from the individual radiant sources. As the number of pixels required to tessellate a given region of the territory increases with decreasing pixel size, the relative performance of the FFT approach in comparison with the traditional pixel-by-pixel calculation procedure will substantially increase for smaller pixels, as indicated by the order of magnitude expressions reproduced after Eq. (10). Note that, in any case, a regular grid is required for applying the FFT approach in Cartesian coordinates.

Let us stress that the particular PSFs [22, 52] used here to exemplify the application of the FFT approach are two among a wide set of possible light pollution propagation kernels. Other particular functions for these or other indicators could be used, accounting for higher-order scattering and for different atmospheric compositions and aerosol concentration profiles. Further developments of this work will use more advanced theoretical models to compute the light propagation kernels. In some cases, experimental PSFs could be determined by observing the light indicator of interest (e.g. the zenithal brightness) at different distances of an effectively small light source, that is, a city, town, or installation whose linear dimensions be small in comparison with the effective height of the atmosphere and the distance to the observation points. Such theoretical or experimental PSFs, adapted to the particular situation addressed by each researcher, can be directly used in the algorithm described here to compute the light pollution effects in wide patches of territory.

Finally let us note that the accuracy of the maps computed using the FFT approach is the same as the one achieved by making the usual pixel-wise calculation, with the difference that the results are obtained in a substantially smaller time. The final accuracy critically depends on the quality of the radiance datasets used to describe the sources, as well as on the appropriateness of the PSF chosen to describe the physics of light propagation in the conditions relevant for the observers. As such, the FFT approach does not improve the physical modeling of the scattering processes, but provides an extremely efficient way of computing their effects, as well as a rich theoretical toolbox for examining them as a particular case of a filtering operation in the framework of two-dimensional linear systems theory [44-45].

## 5. Conclusions

Fast Fourier-transform (FFT) algorithms provide an efficient way of calculating a wide set of magnitudes of interest for light pollution propagation research, with computation times several orders of magnitude smaller than those required by the conventional method of summation over pixels in the spatial domain. The main condition for applying this approach is to reframe the light pollution propagation problem as a convolution integral, i.e. to ensure the spatial shift-invariance of the associated point spread function (PSF).

**Acknowledgments**

This work was supported by Junta de Galicia/FEDER, grant ED431B 2017/64 (SB). CITEUC is funded by National Funds through FCT - Foundation for Science and Technology (project: UID/MULTI/00611/2019) and FEDER - European Regional Development Fund through COMPETE 2020 – Operational Programme Competitiveness and Internationalization (project: POCI-01-0145-FEDER-006922).